\documentclass[aps,prl,twocolumn,10pt,floatfix,longbibliography,superscriptaddress]{revtex4-2}
\usepackage{amsmath,amssymb,bm,graphicx,xcolor,hyperref}
\usepackage[T1]{fontenc}
\usepackage{lmodern}
\hypersetup{colorlinks=true,linkcolor=blue,citecolor=blue,urlcolor=blue}

\usepackage{xcolor}
\usepackage[normalem]{ulem}  
\newcommand{\epsn}{\epsilon_n}
\newcommand{\Vimg}{\mathcal V_n}

\begin{document}
\title{Observation of Scaling of Higher-Order Flow in a Unitary Fermi Gas}

\author{Shuxian Yu}
\affiliation{Institute of Quantum Science and Precision Measurement, School of Physics, East China Normal University, Shanghai 200062, China}
\author{Zhijie Zhang}
\affiliation{Institute of Quantum Science and Precision Measurement, School of Physics, East China Normal University, Shanghai 200062, China}
\author{Yunyi Zhao}
\affiliation{Institute of Quantum Science and Precision Measurement, School of Physics, East China Normal University, Shanghai 200062, China}
\author{Shujin Deng}
\affiliation{Institute of Quantum Science and Precision Measurement, School of Physics, East China Normal University, Shanghai 200062, China}
\author{Meng-Zi Huang}
\affiliation{Institute of Quantum Science and Precision Measurement, School of Physics, East China Normal University, Shanghai 200062, China}

\author{Haibin Wu}
\affiliation{Institute of Quantum Science and Precision Measurement, School of Physics, East China Normal University, Shanghai 200062, China}
\affiliation{Shanghai Branch, Hefei National Laboratory, Shanghai 201315, China}
\affiliation{Collaborative Innovation Center of Extreme Optics, Shanxi University, Taiyuan 030006, China}
\affiliation{Shanghai Research Center for Quantum Sciences, Shanghai 201315, China}

\author{Yu-Gang Ma}
\affiliation{Institute of Quantum Science and Precision Measurement, School of Physics,  East China Normal University, Shanghai 200062, China}
\affiliation{Key Laboratory of Nuclear Physics and Ion-beam Application (MOE), Institute of Modern Physics, Fudan University, Shanghai 200433, China}
\date{\today}

\begin{abstract}

Collective flow emerging from spatially anisotropic many-body systems is a fundamental manifestation of strongly interacting quantum matter, from ultracold quantum gases to the quark-gluon plasma created in relativistic heavy-ion collisions. Yet how collective flow emerges across different spatial harmonics and how dissipation modifies its evolution remain largely unexplored. Here we investigate the nonequilibrium expansion of a unitary Fermi gas initially confined in nearly homogeneous potentials with engineered  angular deformations. By suddenly releasing the confinement, we resolve anisotropic flow harmonics up to the seventh order. Remarkably, after normalization and rescaling by the characteristic expansion timescale, the different harmonics and system sizes exhibit a universal early-time evolution. At longer times, this universality breaks down and we observe an approximately quadratic dependence of the harmonic attenuation on harmonic order, consistent with viscous acoustic scaling. Furthermore, temperature-dependent measurements reveal the role of shear viscosity in the dissipative evolution of higher-order flow. Our results open a new route to exploring the interplay between universal hydrodynamic expansion and dissipative transport, connecting collective dynamics in ultracold quantum gases with anisotropic flow in strongly interacting nuclear matter.
\end{abstract}
\maketitle

Collective flow provides a powerful probe of emergent behavior in strongly interacting many-body systems. In relativistic heavy-ion collisions, the expansion of the hot and dense matter produced by colliding nuclei converts initial spatial anisotropies into correlations in the final momentum distribution \cite{Ollitrault1992}. The resulting azimuthal flow harmonics, conventionally denoted by \(v_n\), have become central observables for understanding the transport properties and collective behavior of the quark--gluon plasma (QGP) \cite{Heinz2013,ShenYan2020,Ma2021,ALICEReview2024}. In particular, elliptic flow (\(v_2\)) and triangular flow (\(v_3\)) encode the response to spatial deformations and their harmonic dependence offers a sensitive probe of dissipative transport \cite{TeaneyYan2011,Schenke2011,Ma2014}, as viscosity preferentially attenuates shorter-wavelength excitations \cite{Deng2024}.
Higher-order harmonics probe smaller spatial structures \cite{STAR2004,ALICE2011,Schenke2012,YanOllitrault2015,ALICEHexagonal2025}; acoustic-scaling analyses organize their relative suppression approximately through $n^2/R$, where $R$ characterizes the initial transverse size \cite{Lacey2014,Liu2018,StaigShuryak2011}.
However, the fluctuating initial geometry and the predominantly final-state access to momentum harmonics make it difficult to resolve how a selected harmonic develops and evolves during expansion \cite{Han2011,Ma2016}.

Ultracold Fermi gases (UFG) offer a unique experimental platform to address such emergent collective flow \cite{Giorgini2008,adams_strongly_2012,Berges2026}. Their initial
density geometry can be precisely engineered, and the temporal evolution of the expansion can be fully reconstructed. Furthermore, the well-characterized equation of state and control of temperature and density provide
independent thermodynamic scales for testing hydrodynamic predictions \cite{Ku2012,Nascimbene2010,Horikoshi2010}. At a Feshbach resonance, the scattering length of two-component Fermi gases diverges, where interactions are as strong as allowed by quantum mechanics (so-called the unitarity limit) and the system becomes scale invariant. Their thermodynamic and transport properties are then governed by universal functions of density and temperature \cite{Enss2011Viscosity,Sommer2011}. Experiments have demonstrated the effects of strong interactions on anisotropic expansion \cite{OHara2002,Clancy2007}, scale invariant dynamics  \cite{Elliott2014,Deng2016Efimovian,Deng2018SuperEfimovian}, shear viscosity \cite{Cao2011,Joseph2015,Bluhm2017}, sound and density response \cite{Baird2019,Patel2020,LiTransport2024,BohlenEnss2020,LiPan2022}. More recently, interaction-induced elliptic flow has been observed in few-fermion systems \cite{Brandstetter2025}, while comparisons between ultracold atomic gases and heavy-ion collisions have revealed common features of anisotropic expansion \cite{Li2025}.

Despite these advances, the dynamical emergence of higher-order flow remains unexplored in ultracold Fermi gases. Most previous expansion studies have focused on global cloud deformation or second-order momentum anisotropy. Higher harmonics are particularly valuable because they probe spatial structures on progressively shorter length scales and therefore provide enhanced sensitivity to viscosity, nonlinear mode coupling and the breakdown of hydrodynamic behavior \cite{Qian2016,Schenke2012}. Moreover, the influence of the initial density distribution can obscure the interpretation of collective flow in conventional harmonic confinement.

\begin{figure*}[htb]
\centering
\includegraphics[width=\textwidth]{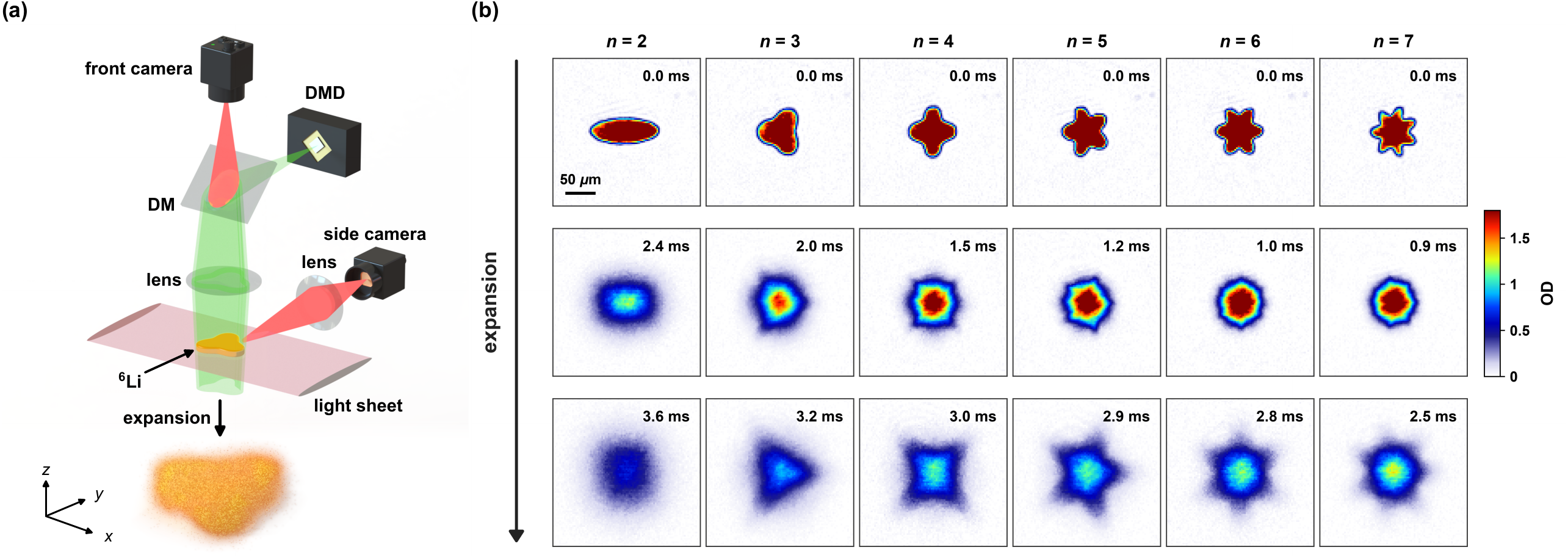}
\caption{\textbf{Programmable generation of higher-order anisotropic flow.} (a) The schematic of the experiment setup. The DMD generates a programmable repulsive hollow beam, preparing a near-uniform $^6$Li cloud in the $x$-$y$ plane. The optical confinement, is released for three-dimensional expansion. Simultaneous front and side imaging provides full information of the expansion and enables thermometry. DM: dichroic mirror. (b) Measured in-situ, intermediate-, and long-TOF density images (top to bottom) for $n=2-7$, colorbar: optical density (OD).}
\label{fig:setup}
\end{figure*}
Here we program angular deformations of nearly homogeneous potentials to investigate the expansion of a unitary Fermi gas. We  measure anisotropic flow harmonics up to the seventh order. We find that, after appropriate normalization, the flow dynamics exhibit a common early-time evolution across different harmonic orders and system sizes. This universal behavior subsequently breaks down, with the different harmonics developing distinct dynamics. We further identify a quadratic harmonic-order dependence of $n^2/R_0$ analogous to acoustic scaling
in nuclear collisions in the measured response and establish a temperature-dependent dissipative contribution to the expansion.
The temperature dependence and time evolution of this scaling provide new constraints on the interplay of pressure-driven conversion and transport. These observations reveal how universal flow generation evolves into harmonic-dependent transport and provide a connection between strongly interacting atomic gases and the acoustic attenuation of anisotropic flow in heavy-ion collisions.

\textit{Experiment.}---We prepare a balanced mixture of the two lowest hyperfine states of $^6$Li close to the broad Feshbach resonance at a magnetic field $B\simeq832\,$G, where the system enters the
unitary regime. A digital micromirror device (DMD) generates a nearly uniform transverse density bounded by an engineered boundary shape [Fig.~\ref{fig:setup}(a)]
\begin{equation}
R(\phi)=R_0+A\cos[n(\phi-\Phi_n)],
\label{eq:boundary}
\end{equation}
where $A$ sets the deformation amplitude and $\Phi_n$ its orientation. We vary $R_0$ and harmonic order $n$ from 3 to 7 while fixing $A/R_0=0.25$. For $n=2$, we prepare a uniform ellipse with a mean radius of $R_0$. Before release, the third direction along $z$-axes is harmonically trapped with frequency $\omega_z/(2\pi)=120(5)\,$Hz. Simultaneous front and side imaging provides thermometry using the equation of state \cite{Ku2012,Elliott2014}. The samples have typical Fermi energy $E_{\rm F}\sim k_B\times 350\,$nK and total atom number $N\sim 1.2\times10^5$ for $R_0\sim30\,\mu$m. We prepare dimensionless temperatures $T/T_{\rm F} \simeq 0.15-0.35$,
where $T_{\rm F}=E_{\rm F}/k_B$.
All optical confinement is suddenly removed, initiating interaction-driven expansion. 
In contrast to harmonically trapped gases, the equilibrium density is nearly uniform and expansion begins through rarefaction waves launched from the boundaries \cite{Wolf2026Rarefaction}, providing a well-defined measurement from spatial deformation to anisotropic collective flow.

The gas evolves for a controlled time of flight (TOF) $t$ at unitarity, and then a standard absorption imaging is performed. The effect from a weak residual magnetic saddle is corrected through a calibrated affine transformation.

For each image we compute the unweighted, signed angular moment of the column density $\Sigma(\bm r,t)$,
\begin{equation}
\Vimg(t)=\frac{\int d^2r\,\Sigma(\bm r,t)\cos[n(\phi-\Psi_n^{\mathrm f})]}{\int d^2r\,\Sigma(\bm r,t)},
\label{eq:observable}
\end{equation}
with coordinates $\bm r$ centered on the cloud and a fixed final-state direction $\Psi_n^{\mathrm f}=\Phi_n+\pi/n$. We use the ratio $\Vimg/\epsn$ to quantify the
efficiency with which the initial spatial anisotropy is converted into collective flow, where
$\epsn=|\langle r^n e^{in\phi}\rangle|/\langle r^n\rangle$
is evaluated from the programmed initial profile, adopting the conventional radial weight in nuclear-flow analyses \cite{Liu2018,TeaneyYan2011}. During early expansion, $\Vimg$ measures spatial anisotropy; after sufficiently long TOF into the ballistic regime, it approaches the corresponding velocity-angle harmonic. At the experimentally accessible TOF, the observed evolution of $\Vimg$ reflects the coupled density and velocity dynamics.

\begin{figure}[tb]
\centering
\includegraphics{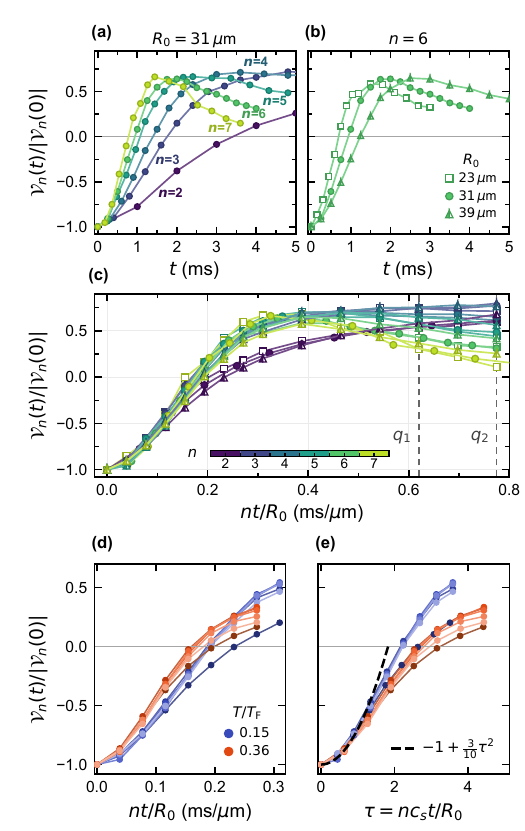}
\caption{\textbf{Universal early-time dynamics of
anisotropic flow.} (a) Normalized anisotropy $\Vimg(t)$ for harmonic orders (colors) of $n=2-7$ versus the TOF at $R_0=31\,\mu$m.  (b) Typical normalized $\Vimg(t)$ of different sizes (markers) for  $n=6$ versus the TOF. (c) Normalized anisotropy $\Vimg(t)$ for all six harmonic orders and three sizes as a function of scaled time $q=nt/R_0$. Dashed vertical lines identify $q_1$ and $q_2$ used in Fig.~\ref{fig:scaling}. (d),(e) Normalized $\Vimg(t)$ for different harmonic orders for two dimensionless temperatures (colors) plotted against $q$ and $\tau=nc_st/R_0$, with $n=2-7$ from dark to light. All early time dynamics collapse with further rescaling by the sound speed. The black dashed curve denotes the quadratic $\tau$ scaling by the  hydrodynamic prediction in Eq.~(\ref{eq:early_quadratic}).}
\label{fig:time}
\end{figure}

\textit{Geometric scaling and universal early-time response.}---We present $\Vimg(t)$ normalized to its initial magnitude in Fig.~\ref{fig:time}(a),(b), showing distinct dynamics for different  $n$ and $R_0$, respectively. The anisotropy begins negative, crosses zero, and subsequently develops resolved maxima for the higher harmonics. At the same temperature $T/T_{\rm F}=0.15(1)$, rescaling the TOF $t$ with $q=nt/R_0$
approximately collapses the early trajectories  and further to the peak time for higher $n>3$ [Fig.~\ref{fig:time}(c)], identifying a geometric timescale $R_0/n$. 

At different temperatures, the trajectories remain distinguishable in $q$ [Fig.~\ref{fig:time}(d)]. 
For nearly homogeneous unitary Fermi gas, the initial density-weighted sound speed is $c_s^2=5\int P_0\,d^3r/(3mN)$ giving an acoustic scale $\tau=c_s q$, where $P_0$ is the initial pressure and $m$ is the atomic mass. When plotted as functions of $\tau$, we observe that the early flow dynamics obtained for different $n$, sizes, and temperatures all collapse onto a common curve [Fig.~\ref{fig:time}(e)], exhibiting the universal early-time response.

This universal behavior can be understood from the boundary driven hydrodynamics of the expansion. Immediately after the
gas is released, the initially near uniform interior remains approximately unperturbed, while rarefaction waves propagate inward from the boundaries. At unitarity, a locally  rarefaction wave follows a self-similar solution of the Euler equations. Its density and velocity profiles depend on distance from the released boundary only through the ratio of that distance to $c_st$.
For zero initial flow and a weak deformation of a uniform transverse disk, $\Vimg(0)\simeq-A/R_0$, $\ddot{\mathcal V}_n(0)/|\Vimg(0)|
\simeq (n^2/R_0^2)\int P_0\,d^3r/(mN)$
and $\dot{\mathcal V}_n(0)=0$, 
this yields
\begin{equation}
\frac{\Vimg(t)}{|\Vimg(0)|}
\simeq -1+\frac{3}{10}\tau^2.
\label{eq:early_quadratic}
\end{equation}
Remarkably, all the rescaled data agree well with this universal prediction up to $\tau\approx 1$. The subsequent separation of the trajectories, including their inversion times, probes the evolving pressure and flow during three-dimensional expansion.

\begin{figure}[t]
\centering
\includegraphics{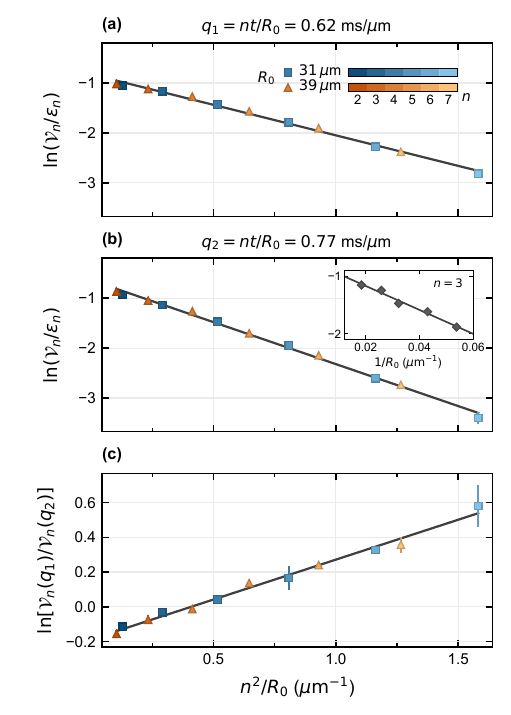}
\caption{\textbf{Quadratic harmonic scaling of late-time anisotropy.} Normalized $\mathcal V_n$ versus $n^2/R_0$ on logarithmic scale at (a) $q_1=0.62\,$ms/$\mu$m and (b) $q_2=0.77\,$ms/$\mu$m. Each panel combines two initial radii (colors) and six harmonic orders (brightness) at $T/T_{\rm F}=0.15(1)$, including the uniform ellipse labeled $n=2$. Black lines are linear fits using Eq.~(\ref{eq:harmonic_scaling}). Inset of (b): varying $R_0$ at fixed $n=3$ tests the radius dependence separately in a wider range. The two-time ratio (c) provides a normalization-independent test of the harmonic scaling. Error bars represent $\pm1$ standard error.} 
\label{fig:scaling}
\end{figure}

\textit{Harmonic scaling at later times.}---At longer expansion times, the normalized flow harmonics depart from their common trajectory. The dynamics become harmonic dependent, revealing a crossover beyond the initial universal regime. Figure~\ref{fig:scaling} compares measurements at two later common rescaled times $q_1<q_2$ as marked in Fig.~\ref{fig:time}(c) at $T/T_{\rm F}=0.15(1)$. Each slice includes two initial radii and harmonic orders up to $n=7$, and is evaluated at $t_j=q_jR_0/n$. Remarkably, the normalized angular moments at these times show an approximately exponential dependence on $n^2/R_0$:
\begin{equation}
\ln\frac{\Vimg(t_j)}{\epsn}
\simeq A_{0,j}-K_j\frac{n^2}{R_0},\qquad j=1,2.
\label{eq:harmonic_scaling}
\end{equation}
Here $K_j$ is an effective attenuation length and $A_{0,j}$ the logarithmic intercept. Linear fits yield $K_1=1.22(3)$\,$\mu$m and $K_2=1.68(4)$\,$\mu$m, with a goodness of fit $R^2$ all above 0.99. The inset of Fig.~\ref{fig:scaling}(b) fixes $n=3$ and varies $R_0$ from 18 to 54~$\mu$m: the logarithmic response again scales approximately linearly with $1/R_0$ at fixed $T_{\rm F}=249(6)$~nK and $T/T_{\rm F}=0.26(7)$. This further confirms the radius dependence over a wider range.

To test this scaling independently of the normalization convention \cite{Han2011,Liu2018}, we plot the two-time logarithmic ratio in Fig.~\ref{fig:scaling}(c),
$\ln[\Vimg(q_1)/\Vimg(q_2)]\simeq-\Delta A_0+\Delta K n^2/R_0$,
where $\Delta A_0=A_{0,2}-A_{0,1}$ and $\Delta K=K_2-K_1$. The $\epsilon_n$ and any time-independent multiplicative normalization cancel. The observed linearity therefore confirms the $n^2/R_0$ dependence of the late-time attenuation independently of the normalization method.

The observed quadratic harmonic order dependence suggests the acoustic attenuation. For a nonrelativistic unitary Fermi gas, the shear-viscous contribution to the sound dispersion is
\begin{equation}
\omega_\pm(k)=\pm c_sk-i\frac{2\eta}{3mn_{0}}k^2+\cdots,
\end{equation}
where \(\eta\) denotes shear viscosity and \(n_0\) is the initial atomic density. The amplitude of a weakly damped sound perturbation therefore decays as $A_k(t)\propto\exp[-Dk^2t]$, where $D=2\eta/3mn_0$. 
For a smooth angular harmonic of order \(n\), the characteristic wave number is approximately $k(t)\simeq n/R(t)$. For a scale invariant expansion, $R(t)^2\simeq R_0^2+v_{\rm exp}^2t^2$ \cite{Elliott2014}, where the expansion speed $v_{\rm exp}$ is on the order of $c_s$. Therefore, the accumulated attenuation exponent at fixed scaled time $q$ is given by $\mathcal{D}_n\sim \int_0^{q R_0/n} dt' Dk^2(t')\sim (n^2D/R_0v_{\rm exp})\arctan(v_{\rm exp}q/n)$. In the late-time limit $v_{\rm exp}q/n\gg1$, this yields the shear-acoustic scaling
\begin{equation}
\mathcal{D}_n \sim \frac{\pi D}{2v_{\rm exp}}\frac{n^2}{R_0}\propto
\frac{\eta}{mn_0c_s}\frac{n^2}{R_0}.
\label{eq:expanding_attenuation}
\end{equation}
Wavelength stretching then makes further attenuation small, so the accumulated $n^2/R_0$ factor persists at later common $q$. The precise value of $\mathcal{D}_n$ is sensitive to the sound diffusivity, the expansion history, and the duration of hydrodynamic evolution. Quantitative comparison with hydrodynamic calculations can disentangle these contributions and constrain transport coefficients.

The $n^2/R_0$ dependence is particularly noteworthy in light of the acoustic scaling of the final momentum anisotropy in relativistic heavy-ion collisions \cite{Lacey2014,Liu2018}. In that setting, the damping of flow
harmonics is governed by the shear viscosity relative to the relativistic entropy density. For a nonrelativistic
Fermi gas, the corresponding inertial quantity is the mass
density $mn_0$. However, the expanding potential differs from a small amplitude sound-wave problem. The flow harmonics are
generated dynamically from an initially isotropic velocity distribution, and the sharp boundaries excite the dynamical expansion with spatial wavelengths. Thus, the observed quadratic scaling has to be understood as an effective collective response rather than a single-mode damping law.

\begin{figure}[t]
\centering
\includegraphics{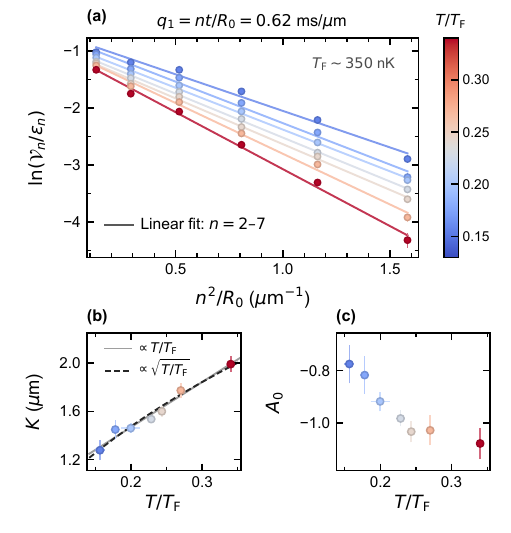}
\caption{\textbf{Temperature dependence of the harmonic attenuation.} (a) $\mathcal V_n/\epsilon_n$ versus $n^2/R_0$ for various dimensionless temperatures at $R_0\simeq31\,\mu$m, with fits at the scaled time $q_1=0.62$~ms/$\mu$m. (b),(c) The slopes $K$ and intercepts $A_{0}$ versus $T/T_{\rm F}$. The gray solid line and the black dashed line in (b) are fits proportional to $T/T_{\rm F}$ and $(T/T_{\rm F})^{1/2}$, respectively, showing that the data are consistent with a range of temperature scaling.}
\label{fig:temperature}
\end{figure}

\textit{Temperature dependence of harmonic attenuation.}---We prepare various temperatures at $R_0\simeq31\,\mu$m for $n=2-7$ at fixed $T_{\rm F}=352(6)$~nK. Figure~\ref{fig:temperature}(a) shows the harmonic scalings at $q_1$ with the fitted slopes and intercepts shown in (b),(c) as a function of the dimensionless temperature. 

The fits proportional to $T/T_{\rm F}$ and $(T/T_{\rm F})^{1/2}$ in Fig.~\ref{fig:temperature}(b) illustrate that our data are consistent with a range of temperature scaling of $K$. The linear dependence, on the one hand, has a simple high-temperature argument from Eq.~(\ref{eq:expanding_attenuation}): when attenuation dominates the harmonic slope, $K$ is proportional to $D/v_{\rm exp}$. At temperatures above the superfluid transition, the gas is a strongly interacting normal quantum fluid. Its shear viscosity can be expressed in a universal form. 
The high temperature unitary-gas transport gives $\eta\propto T^{3/2}$ at fixed density and $c_s\propto T^{1/2}$ \cite{Braby2010,Enss2011Viscosity, Patel2020}, hence
\begin{equation}
K
\propto\frac{1}{k_{\rm F}}\frac{T}{T_{\rm F}},
\label{eq:thermal_benchmark}
\end{equation}
where $k_{\rm F}=(3\pi^2n_0)^{1/3}$ is nearly constant for these preparations. On the other hand, the explored temperature range $T/T_{\rm F}\simeq0.15$--$0.35$ is nevertheless highly degenerate, so Pauli exclusion is expected to play a role that affects the attenuation through both viscosity, expansion speed, and thermal transport. An effective temperature scaling closer to $(T/T_{\rm F})^{1/2}$ in this temperature range is also consistent with previous measurements of the sound diffusivity \cite{Patel2020,LiTransport2024}. Our measurements thus provide new constraints on the interplay of pressure-driven conversion and transport.

\emph{Conclusions.---}
Our measurements revealed the emergence and subsequent evolution of higher-order collective flow in a strongly interacting, scale-invariant, nearly homogeneous quantum fluid.
Programmable angular deformations revealed an early universal acoustic response and later $n^2/R_0$ scaling for harmonic orders up to $n=7$. 
Ideal hydrodynamics quantitatively explained the initial quadratic evolution, while viscous acoustic attenuation captured the late-time harmonic scaling and its temperature dependence.

Looking forward, harmonic-resolved expansion measurements could provide access to temperature dependent transport coefficients to test string theory limit \cite{KSS2005,adams_strongly_2012} and nonlinear flow couplings, including the conversion of lower-order modes into higher harmonics. The approach developed here could also be straightforwardly generalized across the entire BEC–BCS crossover regime, enabling the investigation of how changes in interaction strength and scale invariance affect the emergence of collective flow. 
This work thereby extends quantum gas experiments from elliptic flow to higher-order anisotropic flows, opening up new opportunities to connect programmable ultracold quantum gases with collective dynamics in strongly interacting nuclear matter.

\begin{acknowledgments}
\emph{Acknowledgments.---}
The project was supported by National Key R\&D Program of China (No.~2022YFA1404202), National Science Foundation of China (No.~12234008, No.~12688101, No.~12547102), the Shanghai Municipal Science and Technology Major Project (No.~2019SHZDZX01, No.~25LZ2600400), Innovation Program for Quantum Science and Technology (No.~2021ZD0302100), and Shanghai QiYuan Innovation Foundation.
\end{acknowledgments}
\bibliography{references}
\end{document}